\documentclass[11pt,twoside]{article}

\usepackage{asp2006}
\usepackage{epsf}
\usepackage{psfig}
\usepackage{lscape}
\usepackage{graphicx}

\begin{document}
\title{Multi-ring structure of the eclipsing disk in EE~Cep - possible
planets?}

\author{C.~Ga\l{}an,$^1$ M.~Miko\l{}ajewski,$^1$ T.~Tomov,$^1$
E.~\'Swierczy\'nski,$^1$ M.~{Wi\c{e}cek},$^1$ T.~Bro\.zek,$^1$
G.~Maciejewski,$^1$ P.~Wychudzki,$^1$ M.~Hajduk,$^1$
P.T.~R\'o\.za\'nski,$^1$ E.~Ragan,$^1$ B.~Budzisz,$^1$ P.~Dobierski,$^1$
S.~{Fr\c{a}ckowiak},$^1$ M.~Kurpi\'nska-Winiarska,$^2$ M.~Winiarski,$^{2,3}$
S.~Zo\l{}a,$^{2,3}$ W.~Og\l{}oza,$^3$ A.~Ku\'zmicz,$^2$ M.~Dr\'o\.zd\.z,$^3$
E.~Kuligowska,$^2$ J.~Krzesi\'nski,$^3$ T.~Szyma\'nski,$^2$ M.~Siwak,$^4$
T.~Kundera,$^2$ B.~Staels,$^5$ J.~Hopkins,$^6$ J.~Pye,$^7$ L.~Elder,$^7$
G.~Myers,$^8$ D.~Dimitrov,$^9$ V.~Popov,$^9$ E.~Semkov,$^9$ S.~Peneva,$^9$
D.~Kolev,$^{10}$ I.~Iliev,$^{10}$ I.~Barzova,$^{10}$ I.~Stateva,$^9$
N.~Tomov,$^{10}$ S.~Dvorak,$^{11}$ I.~Miller,$^{12,13}$ L.~Br\'at,$^{14,15}$
P.~Niarchos,$^{16}$ A.~Liakos,$^{16}$ K.~Gazeas,$^{17}$ A.~Pigulski,$^{18}$
G.~Kopacki,$^{18}$ A.~Narwid,$^{18}$ A.~Majewska,$^{18}$
M.~{St\c{e}\'slicki},$^{18}$ E.~Niemczura,$^{18}$ Y.~\"O\v{g}men,$^{19}$
A.~Oksanen,$^{20}$ H.~Ku\v{c}\'akov\'a,$^{21}$ T.A.~Lister,$^{22}$
T.A.~Heras,$^{23}$ A.~Dapergolas,$^{24}$ I.~Bellas-Velidis,$^{24}$
R.~Koci\'an,$^{21}$ A.~Majcher$^{25}$ }

\affil{
\footnotesize
$^1$Nicolaus Copernicus University, ul. Gagarina 11, 87-100 Toru\'n, Poland\\
$^2$Astronomical Observatory, Jagiellonian Univ., ul. Orla 171, 30-244 Krak\'ow, Poland\\
$^3$Mt. Suhora Obs., Pedagogical Univ., ul. {Podchor\c{a}\.zych} 2, 30-084 Krak\'ow, Poland\\
$^4$DAA, University of Toronto, 50 St.George St., M5S 3H4 Toronto, Ontario\\
$^5$Sonoita Research Observatory/AAVSO\\
$^6$Hopkins Phoenix Obs., 7812 West Clayton Drive, Phoenix, Arizona 85033-2439 USA\\
$^7$Maui Community College, Kahului, Hawaii\\
$^8$GRAS Observatory, Mayhill, New Mexico\\
$^9$Institute of Astronomy, BAS, 72, Tsarigradsko Shose blvd., BG-1784 Sofia, Bulgaria\\
$^{10}$NAO Rozhen, Institute of Astronomy, BAS, PO Box 136, 4700 Smolyan, Bulgaria\\
$^{11}$Rolling Hills Observatory Clermont, FL USA\\
$^{12}$Furzehill House, Ilston, Swansea. SA2 7LE. UK\\
$^{13}$Variable Star Section of the British Astronomical Association\\
$^{14}$Variable Star and Exoplanet Section of Czech Astronomical Society\\
$^{15}$Altan Observatory, Velka Upa 193, Pec pod Snezkou, Czech Republic\\
$^{16}$Department of Astrophysics, Astronomy and Mechanics, National and Kapodistrian University of Athens, GR 157 84 Zografos, Athens, Greece\\
$^{17}$Harvard-Smithsonian Center for Astrophysics, 60 Garden St., Cambridge, MA 02138, USA\\
$^{18}$Astronomical Institute, Wroc\l{}aw Univ., ul. Kopernika 11, 51-622 Wroc\l{}aw, Poland\\
$^{19}$Green Island Observatory (B34), North Cyprus\\
$^{20}$Hankasalmi~Obs.,~Jyvaskylan~Sirius~ry,~Vertaalantie~419,~FI-40270~Palokka,~Finland\\
$^{21}$The Observatory and Planetarium of Johann Palisa VSB, - Technical University of Ostrava, 17. listopadu 15, 708 33 Ostrava-Poruba, Czech Republic\\
$^{22}$Las Cumbres Observatory, 6740 Cortona Drive Suite 102, Goleta, CA 93117, USA\\
$^{23}$Observatorio Astron\'omico "Las Pegueras", NAVAS DE ORO (Segovia), Spain\\
$^{24}$Institute~of~Astronomy~and~Astrophysics,~NOA,~PO~Box~20048,~11810~Athens,~Greece\\
$^{25}$Soltan Institute for Nuclear Studies, Warsaw, Poland
}

\begin{abstract}
The photometric and spectroscopic observational campaign organized for the
2008/9 eclipse of EE~Cep revealed features, which indicate that the
eclipsing disk in the EE~Cep system has a multi-ring structure.  We suggest
that the gaps in the disk can be related to the possible planet formation.
\end{abstract}

\section{Introduction}

The eclipses of the 11th magnitude star EE Cep have been observed with
a~period of 5.6 yr from the early 1950-ies.  Their depths change in a wide
range from about $0\fm5$ to $2\fm0$ \citep[see][]{Gra2003}, however all of
them show the same features: they are almost gray and have the same
asymmetric shape (descending branch of every eclipse has longer duration
than the ascending one).  In all eclipses it is possible to distinguish five
phases (shown in Fig.~1): ingress (1--2) and egress (3--4), respectively
preceded and followed by the extended atmospheric parts (1$_{\rm a}$--1 and
4--4$_{\rm a}$), and the slope-bottom transit (2--3).  The most plausible
explanation of the observed photometric behavior was proposed by
\citet{MiGr1999}.  They suggested that the secondary consists of a dark
disk, with opaque interior and semi-transparent exterior, around a low
luminosity central object.  The inclination of the disk to the line of sight
and the tilt of its cross-section to the direction of motion are changed by
precession.  This causes the changes in the depth and the duration of the
eclipses.  A significant impact parameter is responsible for the observed
asymmetry of the eclipses.  This model can explain the shallow ($0\fm6$),
flat-bottomed eclipse observed in 1969 if we assume a nearly edge-on and
non-tilted projection~of~the~disk.

\begin{figure}[!ht]
\begin{center}
\includegraphics[angle=0, width=0.64 \textwidth]{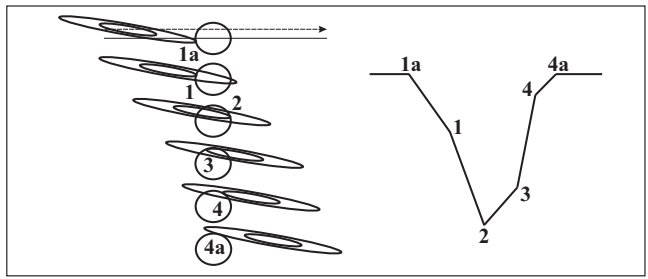}
\end{center}
\caption{\small A schematic illustration of the eclipses in the EE~Cep
system.  The contact times ($1_{\rm a},1,2,3,4,4_{\rm a}$) distinguished in
the light curve (right) refer to characteristic configuration of the disk
and the star during~the~eclipse~(left).}
\end{figure}

\section{The 2003 eclipse}

An observational campaign organized during the 2003 eclipse \citep{Mik2003},
brought very good quality photometric $UBV(RI)_{\rm C}$ data with a dense
time coverage.  For the first time it was possible to analyze the color
evolution during the eclipse and not only their amplitudes.  The preliminary
photometric results of this campaign were described by \citet{Mik2005a}. 
The eclipse turned out to be quite shallow and in accordance with
expectations almost gray.  The eclipse achieved depths from about $0\fm5$ in
$I_{\rm C}$ to $0\fm7$ in $U$ passbands.  In Fig.~2, the $B$ light curve
and the color indices are presented.  Each point represents the average of
all measurements obtained in a given passband during a single night.  The
color indices from the 2003 eclipse show two blue maxima, about nine days
before and after the mid-eclipse.  Simultaneously, weak maxima in the $B$
light curve are clearly visible.

\begin{figure}[!ht]
\begin{center}
\includegraphics[angle=0, width=0.63 \textwidth]{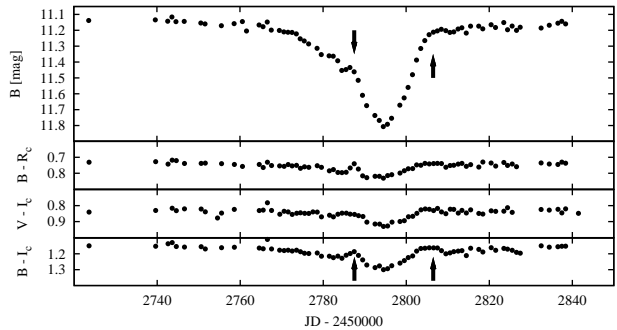}
\end{center}
\caption{\small The 2003 eclipse. The $B$ light curve (top) and three color
 indices (bottom) are presented. Arrows denote times of blue maxima.}
\end{figure}

\section{The gapped disk model}

The blue maxima observed in the color indices can be understood if we assume
that a hot star, rotationally darkened at the equator and brightened at the
poles, is eclipsed by a disk divided into two parts by a gap (Fig.~3). 
The spectra obtained in 2003 indicate indeed that the hot component
is~a~rapidly rotating Be star \citep{Mik2005b}.  A comparison of the Balmer
absorption lines in the spectrum of EE Cep with theoretical profiles gives
$v \sin{i} \approx 350$ km/s \citep[see][]{Gal2008}, which implies strong
equatorial darkening.  The difference between the polar and the equatorial
temperatures can reach even~5-6~$\cdot 10^3$~K.

\begin{figure}[!ht]
\begin{center}
\includegraphics[angle=0, width=0.75 \textwidth]{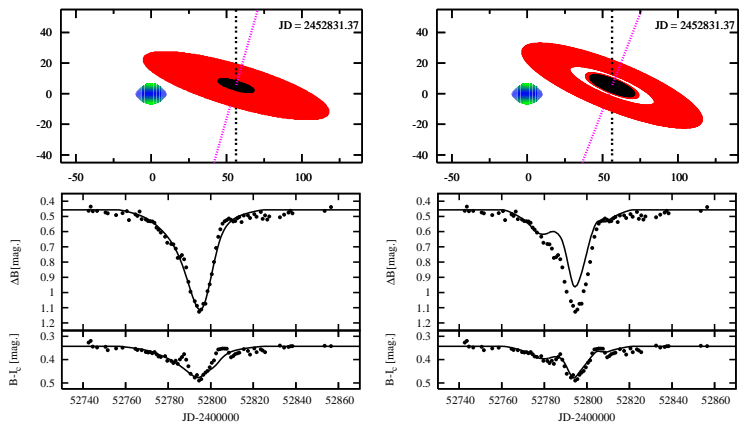}
\end{center}
\caption{\small Modeling the eclipse of a fast rotating Be star by a solid
disk~(left) and by a disk with a gap (right).  A flat, circular disk with
the $r^{-2}$ density profile has been assumed.  {\textit {Top:}} the system
projection in the plane of the sky.  The polar~(hot) and the equatorial
(cool) areas of the star are shown by changing shades.  The inner (opaque)
and the outer (semi-transparent) areas of the disk are shown by dark and
light shades, respectively.  The sizes are expressed in the solar radii. 
{\textit {Middle:}} The $B$ light curve (points) from Miko\l{}ajewski et al. 
(2005a) fitted with synthetic curves (lines).  {\textit {Bottom:}} The
$B-I_{\rm C}$ color index from the 2003 eclipse (points) together
with~the~synthetic~fits~(lines).}
\end{figure}

\noindent Therefore, when the hot polar area appears in the gap the blue
maxima are observed.  In Fig.~3 two eclipse models are presented: with
a~solid disk and with a disk having a~gap.  The first model gives quite
a~good fit to the light curve and the global color changes, but it does not
explain in details the color evolution during the eclipse.  The disk model
with a~circular gap fits the observed color changes quite well.  It was not
possible to obtain a~good fit to the light and color curves simultaneously,
most likely because the assumed formula for the density of the disk was too
simple.

\section{The 2008/9 eclipse}

The fruitful observational campaign in 2003 did not provide answers to a
number of questions: (i) is the Be star (primary) in the EE~Cep system
indeed eclipsed by a dark, precessing disk?  (ii) does the gap in the disk
really exist?  (iii) what is the nature of the central body surrounded by
the disk and what is its contribution to the total flux?  An excellent
opportunity for answering these questions came with the next eclipse, which
took place at the turn of 2008, with a minimum occuring on January 10, 2009. 
The invitation to the observational campaign was announced in a short paper
in {\textit {IBVS}} \citep{Gal2008}.

\begin{figure}[!ht]
\begin{center}
\includegraphics[angle=0, width=0.97 \textwidth]{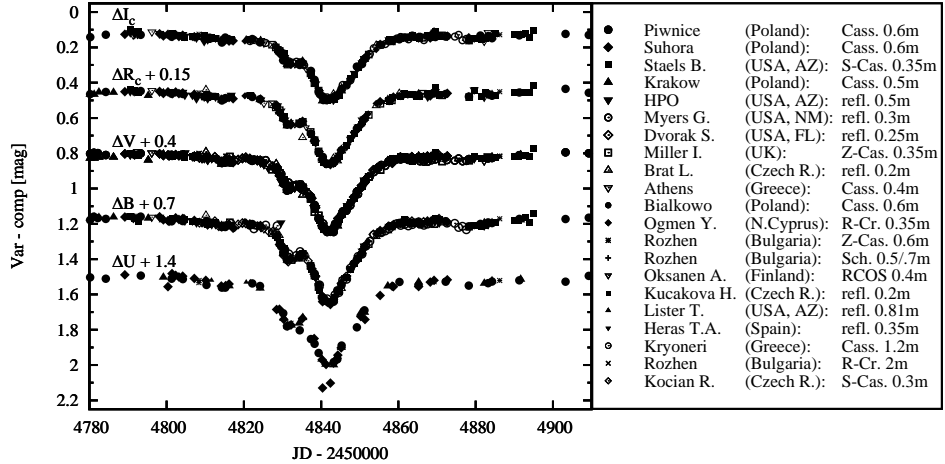}
\end{center}
\caption{\small The $UBV(RI)_{\rm C}$ photometry obtained during 2008/9
eclipse using 21 telescopes located in Europe and North America.}
\end{figure}

The strong interest among observers, who took part in the observations,
resulted in a large amount of the collected data.  The $UBV(RI)_{\rm C}$
light curves presented in Fig.~4 are composed of more than 1600 individual
data points in total!  Surprisingly, the last eclipse with depths
$\sim0\fm5$ in $U$ and nearly $\sim0\fm4$ in $I_{\rm C}$ passbands, turned
out to be the shallowest in the observing history of EE~Cep.  In Fig.~5, we
present the {\it B} light curve and the color indices.  Each point
represents the average of all measurements obtained in a given filter during
the first and second part of a particular Julian day.  The accuracy of the
photometry is excellent, reaching a few mmag.  The features observed during
the previous eclipse, including the two blue maxima, occur also in the
2008/9 eclipse.

\begin{figure}[!ht]
\begin{center}
\includegraphics[angle=0, width=0.64 \textwidth]{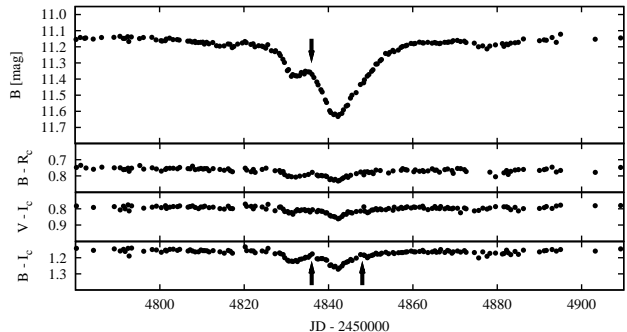}
\end{center}
\caption{\small The 2008/9 eclipse. The $B$ light curve (top) and three
color indices (bottom) are presented. Arrows denote times of blue maxima.}
\end{figure}

\noindent The bump preceding the minimum, at JD~2454836 (Fig.~4~and~5) is
much more pronounced than previously.  The differences in the phase and
strength of these features can be caused by the changes in the spatial
orientation of the disk.  The observed variations in the {\it I} passband
before and after the eclipses during the last decade \citep{Gal2008} would
give additional support for this idea.

\section{Conclusions}

A comparison of the photometric and spectroscopic data obtained during the
last two eclipses (see Fig.~6) reveals some new characteristics of the
eclipsing disk in the EE~Cep system.  The durations of last two eclipses
were longer than we expected (about 90 days), and they both were preceeded
and followed by very shallow minima which are perhaps repeatable in each
orbital cycle.  The 2008/9 campaign results confirmed the existence of a gap
in the disk.  Additionaly, the data present some indications of the
existence of a~second, outer gap in the disk.  The possible multi-ring
structure of the eclipsing disk in EE~Cep suggests the existence of some
massive bodies that could be responsible for their formation.  This means
that we may observe signs of planet formation in a circumstellar disk in
EE~Cep.  The results presented here show that the disk in EE~Cep system
is very similar to the multi-ring structure observed in $\varepsilon$
Aurigae \citep{Fer1990}.

During the EE~Cep eclipses additional components appear in the $H_{\alpha}$
emission line and in the NaI absorption doublet.  Towards the mid-eclipse an
absorption component appears and grows in the $H_{\alpha}$ profile and
during the minimum it is very deep and broad.  The sodium doublet line
profiles evolve during the eclipse and in the minimum multi-component
structure with at least two additional absorption components can be seen. 
During the shallow minimum at orbital phase 0.017 an additional NaI
absorption component was also present, while it was absent soon after the
egress.  The spectra from the last two eclipses suggest that the absorption
lines evolution is the same during each cycle.  Unfortunately spectroscopic
observations coverage of the last eclipse was very bad.  Many more
spectroscopic observations during the next eclipse would be needed to
understand the nature of EE~Cep.  Photometry in the infrared {\it JHK}
passbands and the radial velocity variations of the hot component
could~be~invaluable~as~well.

\begin{figure}[!ht]
\begin{center}
\includegraphics[angle=0, width=0.98 \textwidth]{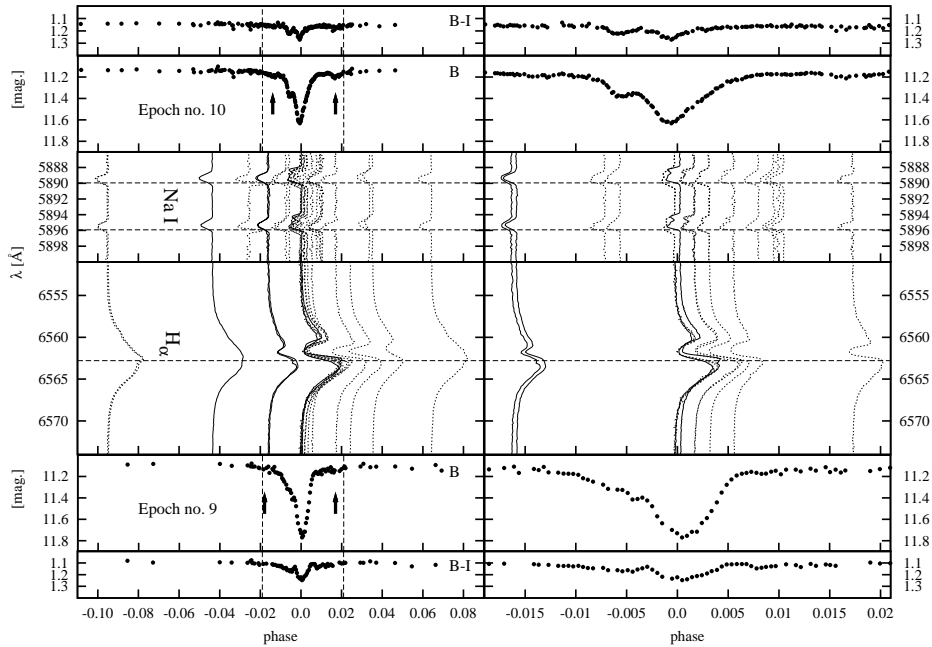}
\end{center}
\caption{\small Photometry and spectra from last two eclipses are compared. 
{\textit {Left:}} The observations between orbital phase 0.9 and 0.1 are
presented.  {\textit {Right:}} An expanded view of 0.04 of orbital phase
marked with vertical dashed lines in the left panel is shown.  {\textit
{Top:}} Photometry of the 2008/9 eclipse.  {\textit {Bottom:}} Photometry of
the 2003 eclipse.  The times close to the very shallow minima which occur in
both eclipses at phase near of $\pm0\fp017$ are denoted with arrows. 
{\textit {Middle:}} The evolution of the NaI absorption and the H$_{\alpha}$
emission lines.  The spectra obtained during and close to the 2003 eclipse
are given with dashed lines, while those obtained during the last eclipse,
with solid lines.  The positions of the continuum levels refer to the
orbital phases.}
\end{figure}

\acknowledgements A part of the observations used were taken through
courtesy of the AAVSO and the Sonoita Research Observatory.  This study was
supported by MNiSW grant No.~N203~018~32/2338 and UMK promotor's grants No. 
366-A and 367-A.

\noindent {\textit{Devinney:}} Do you have a mass function, $f(m_{\rm 2})$, for the B5III
component?\\

\noindent {\textit {Ga\l{}an:}} No. We have not spectroscopic orbit.

\end{document}